\documentclass[12pt]{article}
\usepackage{amssymb,amsmath,graphicx}
\usepackage{verbatim} 

\begin{document}
\begin{titlepage}
\begin{center}
{\Large \bf Dynamical Decompactification and Three Large Dimensions}

\vspace{6mm}
\renewcommand\thefootnote{\mbox{$\fnsymbol{footnote}$}}
Brian Greene${}^1$\footnote{greene@physics.columbia.edu},
Daniel Kabat${}^{1,2}$\footnote{daniel.kabat@lehman.cuny.edu} and
Stefanos Marnerides${}^1$\footnote{stefanos@phys.columbia.edu}

\vspace{4mm}

${}^1${\small \sl Institute for Strings, Cosmology and Astroparticle Physics} \\
{\small \sl and Department of Physics} \\
{\small \sl Columbia University, New York, NY 10027 USA}

\vspace{4mm}

${}^2${\small \sl Department of Physics and Astronomy} \\
{\small \sl Lehman College, City University of New York} \\
{\small \sl Bronx, NY 10468 USA}

\end{center}

\vspace{1cm}

\noindent
We study string gas dynamics in the early universe and seek to realize the
Brandenberger - Vafa mechanism -- a goal that has eluded earlier works -- that singles out three or fewer spatial dimensions as the number which grow large cosmologically.
Considering wound string interactions in an impact parameter picture, we show that
a strong exponential suppression in the
interaction rates for $d>3$ spatial dimensions reflects the classical
argument that string
worldsheets generically intersect in at most four spacetime dimensions.
This description is appropriate in the early universe if wound strings are
heavy -- wrapping long cycles -- and \textit{diluted}. We consider the
dynamics of a string gas coupled to dilaton-gravity and find that
a) for any number of dimensions the universe generically stays trapped in the
Hagedorn regime and b) if the universe fluctuates to
a radiation regime any residual winding modes are diluted enough so that they freeze-out in $d>3$ large
dimensions while they generically annihilate for $d=3$. In this sense the
Brandenberger-Vafa mechanism is operative.   

\end{titlepage}
\setcounter{footnote}{0}
\renewcommand\thefootnote{\mbox{\arabic{footnote}}}

\section{Introduction}

One of the few mechanisms aiming to explain the hierarchy between
three large and six small spatial dimensions within superstring theory
is due to a suggestion, some two decades ago, by  Brandenberger and Vafa \cite{BV}; see \cite{wats} for a
review. In this scenario the early universe consists of a hot string
gas in thermal equilibrium near the Hagedorn temperature. The topology
of space has non-trivial cycles supporting winding modes in the
gas. The background metric and string coupling evolve with the low
energy effective dilaton-gravity equations of motion according to
which the winding modes resist the expansion of the spatial directions
they wrap. If due to a thermal fluctuation a number of dimensions
starts growing then eventually the equilibrium number of winding modes
will drop to zero. The winding modes have the capacity to relax to equilibrium through
annihilations with anti-winding
modes;  if these interactions are efficient then at large volumes the
winding numbers will vanish allowing the corresponding dimensions to
grow. The Brandenberger-Vafa (BV) mechanism relied on a simple
dimension-counting argument that wound strings generically intersect
in at most 3 spatial dimensions, singling this out as the maximum number of
dimensions in which winding numbers have the capacity to track their equilibrium values, thereby dropping to zero and allowing the dimensions to grow large. In \cite{sak} this argument was
supported using numerical simulations of a network of classical
strings, though gravitational dynamics was not taken into account.

Over time it became clear that strings are not the only fundamental
degrees of freedom of string theory and that higher dimensional
objects (membranes) are also fundamental states of the
theory; superstring theory was shown to result from
compactification of a higher dimensional theory, M-theory. In a paper
by Alexander, Brandenberger and Easson \cite{ABE} the setup of
\cite{BV} was extended to include p-branes for $p=0,1,2,4,5,6,8$ in
the weak-coupling limit of M-theory with one small dimension
compactified on $S^1$ (type IIA string theory). The other spatial
dimensions were compactified on a 9-torus. The authors argued that
fundamental string winding modes are still the decisive objects
regarding decompactification and that the conclusions of \cite{BV}
still hold. They also pointed out that a further hierarchy between
dimensions could arise. Past the string scale, as the universe grows,
more and more energy is needed to support wound branes of highest $p$,
hence highest $p$ branes would tend to decay first. As two $p$-branes
can intersect in at most $2p+1$ spatial dimensions, there is no
obstacle for the disappearance of $p$-branes for $p>2$. But 2-branes
can allow for a 5-dimensional subspace to grow first. Further, within
this subspace, 1-branes will only allow for a 3-dimensional space to
continue expanding, as in \cite{BV}, hence one is left with a 3-2-4
dimensional hierarchy.

These claims relied on heuristic thermodynamic and topological
arguments.  Aiming to carry out a more rigorous investigation, Easther
et al.\ \cite{late-branes} considered the full equations of motion for
11D supergravity on a homogeneous but anisotropic toroidal background,
coupled to a gas of branes and supergravity particles. Focusing on the
late time behavior of the system, they justifiably ignored excitations on the
branes and included only M2-branes, since M5-branes (the other
fundamental states of M-theory) would annihilate efficiently in the
full 11-dimensional spacetime. Motivated by the BV mechanism and the
arguments of \cite{ABE}, the authors of \cite{late-branes} chose
initial states resulting from fluctuations that would leave 3
dimensions unwrapped, some number of dimensions partially wrapped and
some fully wrapped. The conclusion was that indeed the dynamics
leverage the topological reasoning and a hierarchy among dimensions is
established. This conclusion was supported further by \cite{campos},
where in addition nontrivial fluxes were included (the 3-form gauge
field of 11D supergravity). In fact, the presence of fluxes seemed to
enlarge the possible space of initial conditions that lead to three
large dimensions at late times. Specifically, for the case of six
initially unwrapped dimensions, the dynamics of fluxes introduced a
new hierarchy suppressing the growth of 3 out of the 6 unwrapped
dimensions.

An apparent limitation of the BV argument is that it seems to depend crucially on non-contractible spatial
cycles and their associated topologically stable winding modes.
Phenomenologically viable compactifications of string theory, however, may not
have such cycles.  Nonetheless the authors of \cite{orb} surmised that these more general spaces might still
support ``pseudo-wound'' modes, long strings that extend around a
dimension but are contractible. If these strings are stable over time scales
larger than the cosmological Hubble scale, then as far as the dynamics
are concerned they play the same role as stable wound strings. In
\cite{orb}, using numerical simulations for string networks on
toroidal orbifolds with trivial fundamental group, the authors showed
that pseudo-wound strings generically do persist for many Hubble
times, suggesting that the requirement of non-contractible cycles can
be relaxed.

The results up to this point seemed promising, but it remained to
actually test the heart of the argument: whether at early times,
thermal fluctuations near the Hagedorn era and string (or brane)
interactions really lead to annihilation of winding modes in a
3-dimensional subspace. An early attempt to investigate this was
carried out in \cite{early-branes}. The authors considered a gas of
2-branes and supergravity particles, along with excitations on the
branes that lead to a limiting Hagedorn temperature. This setup was
within the low-energy limit of M-theory compactified on a 10-torus,
with an anisotropic and homogeneous metric evolving according to
11-dimensional Einstein gravity.  The winding numbers of 2-branes
evolved according to Boltzmann equations. The authors assumed initial
conditions in which the total volume of the torus was fixed but
otherwise assumed that all states were equally likely. By numerically
solving the coupled Boltzmann-gravity equations the authors concluded
that the number of unwrapped dimensions at late times depended
crucially on the initial volume of the torus. Typically a large (and
monotonically increasing) overall volume would decrease the
interaction cross-section of branes too quickly, eventually leading to
brane number freeze-out. If the initial volume was
constrained according to holographic arguments, the initial winding
numbers proved so small that all dimensions would decompactify early
on. Three dimensions was not found to be singled out by the dynamics.

Similar all-or-nothing behaviour was found in
\cite{winds,Danos:2004jz} for IIA theory compactified on $T^9$.  In
these papers the dilaton-gravity equations for the background were
coupled to Boltzmann equations for winding modes and radiation.
Even though this behaviour was attributed to the rolling of the
dimensionally reduced coupling to weaker values, we emphasize here a
more decisive phenomenon that prevents the annihilation of the winding modes and
yields the all-or-nothing behaviour. If the initial energy density of the
universe is large, the system is found in the Hagedorn phase with a significant
amount of winding present in thermal abundance but in a regime that resembles
a matter-dominated universe with vanishing pressure. This, along with
``friction'' due to the dilaton's velocity, results in an insignificant growth
of the wrapped dimensions (even over an infinite amount of time).  With the
total energy nearly constant, the equilibrium number
of winding modes generically does not fall to zero. In this sense the system stays ``trapped'' in the Hagedorn phase.
It is very likely, however, that this problem is particular to the
approximation of treating the background with the lowest order dilaton-gravity
dynamics.  Corrections to these (\cite{cyc}), or a different
treatment of the metric degrees of freedom, could alleviate it. An alternative
approach would be to keep the background dynamics to lowest order and still consider a high density initial phase -- a fairly
natural assumption -- but consider large volume fluctuations that could yield an exit.
This is the approach we adopt here.

Finally, and most importantly for our purposes, the aforementioned problems were 
independent of the number of dimensions growing large. The reason
was that the rate at which
wound strings annihilated only fell off like the inverse volume of the
transverse dimensions.  This failed to single out three large
dimensions as special, suggesting that the Brandenberger-Vafa argument might not be supported by
the dynamics underlying string/M-theory.

In this paper, we re-examine this conclusion and suggest a possible way in which string dynamics
may indeed favor three large dimensions. Our basic approach is this:
According to the Brandenberger-Vafa dimension-counting argument,
one expects that string interaction rates should be dramatically
suppressed when the number of large spatial dimensions is bigger than
three. Moreover, as the dimension-counting argument is purely classical, one
expects it to be valid in a regime where the wound strings behave
nearly classically and can be regarded as one-dimensional extended
objects tracing a two-dimensional worldvolume. In such a regime the
quantum thickness of the strings should be small compared to their
length along the dimension they wrap and also small compared to the
size of the transverse space. This suggests, in contrast to our previous
work, that we hope for a {\it dilute} gas of winding strings.  
Furthermore, as we discuss, in a dilute regime we are lead to work in an impact
parameter representation of the string scattering amplitude. As we will see, this
makes manifest the distinction between three and more large
spatial dimensions regarding the interactions of winding modes. 
Our main observation is that if the universe fluctuates out of an initial dense
Hagedorn regime -- something that we believe is generically necessary in order
to match to a realistic expanding cosmology in the string gas scenario
independently of the decompactification mechanism -- then any residual winding
modes that were thermally excited in the Hagedorn phase, are indeed diluted
enough so that they freeze-out in $d>3$ spatial dimensions. Further, for
$d=3$, the enhancement in interaction rates due to the length of wound strings
generically overcomes the suppression due to the weak coupling and winding
modes may annihilate efficiently. 
While a number of important issues remain,
this appears to be the first demonstration of dynamical string theory decompactification that generically yields three large spatial dimensions.

By way of outline we begin with a discussion of the impact parameter
representation, proceed to set up our model for the string gas, and
finish with a numerical simulation, along the lines of \cite{winds},
that will allow us to identify the regions of phase space in which
three or more spatial dimensions decompactify.

\section{Interaction amplitudes and impact parameter picture}
In this section we derive interaction rates for wound strings in a
semiclassical impact parameter picture. We will show that when long
strings interact at impact parameters larger than their thickness,
there is an exponential suppression in the interaction rates for $d>3$.

The starting point is the Virasoro-Shapiro amplitude for wound strings in $d=D-1$
large dimensions given by \cite{polch2,polch1}
\begin{equation}\label{amp}
A(s,t)=-\kappa_{D-2}^2\frac{s^2}{t}(\alpha' s/4)^{\alpha't/2}e^{-i\pi\alpha't/4} 
\end{equation}
with $s$ computed either from the right-moving or left-moving momenta of the
closed string, $s \approx 4 R^2 / \alpha'^2$ with $R$ the radius of the
dimension that
the strings wrap. The imaginary part of the amplitude as $t\rightarrow 0$ is
\begin{equation}\text{Im} \, A(s,t=0) = \frac{\alpha'\pi}{4}\kappa_{D-2}^2s^2\end{equation}
Here $\kappa^2_{D-2} = \kappa^2 / V$ is the gravitational coupling in
$D-2$ dimensions, where $V$ is the transverse compactification volume
times the area of the torus wrapped by the strings \cite{polch2}.  By
the optical theorem $\frac{1}{s}\text{Im}(A(s,t))|_{t=0} \sim
\alpha'\kappa_{D-2}^2s$ controls string interactions. It is crucial to observe
that for $D=4$ this
quantity is dimensionless and gives the probability for two colliding
winding strings to interconnect and unwind (to leading order), while for $D>4$ it has
units of (length)$^{D-4}$ and represents a cross section in the $D-4$
dimensions transverse to the moving strings. This reflects the fact
that long strings generically intersect in $D=4$, like point particles
moving on a line, while they generically miss in $D>4$ and the
relevant quantity becomes a cross-section.

One can consider the interaction probability in an impact
parameter picture. As discussed above, long wound strings have an
effective impact parameter in the $D-4$ directions transverse to the
motion of both strings. The impact parameter $b$ is the conjugate
variable to the transverse momentum $q=\sqrt{-t}$ and the amplitude in
this representation is obtained by the following transform in the
transverse directions.
\begin{equation}\label{transf}
A(s,b)=\int \frac{d^{D-4}q}{(2\pi)^{D-4}}e^{-iqb} \frac{A(s,t)}{s}
\end{equation}
For the Virasoro-Shapiro amplitude (\ref{amp}), using
$q^{-2}=\alpha'\int^1_0dx\,x^{\alpha'q^2-1}$, this gives
\begin{equation} 
\begin{aligned}
A(s,b) &= \alpha'\kappa_{D-2}^2s\int^1_0\frac{dx}{x} \int \frac{d^{D-4}q}{(2\pi)^{D-4}}e^{-(Y-i\frac{\pi}{4}-\log(x))\alpha'q^2-ibq}\\
&=\frac{\kappa_{D-2}^2s}{4\pi^{(D/2)-2}}b^{6-D}\mathcal{\gamma}\Big(\frac{D}{2}-3,\frac{b^2/(4\alpha')}{Y-i\frac{\pi}{4}}\Big)
\end{aligned}
\end{equation}
where $Y=\log(\frac{\alpha's}{4})$ and $\gamma(a,x)$ is the lower
incomplete gamma function. The imaginary part of the above amplitude
in the limit $b^2\gg Y\alpha'$ is
\begin{equation}\label{im}
\text{Im} \, A(s,b) \rightarrow \frac{\pi\alpha'\kappa_{D-2}^2s}
{4(4\pi Y\alpha')^{D/2-2}}e^{-\frac{b^2}{4Y\alpha'}}
\end{equation}

These results are similar to those found in \cite{ven}, the difference being
that the authors of \cite{ven} consider graviton scattering and take
the number of transverse directions to be $D-2$. In fact the
interpretation of $b$ as a classical impact parameter in
(\ref{transf}) can be justified along the lines of \cite{ven}.  In the
high energy limit ($s\rightarrow\infty$ which for wound strings is
$R\rightarrow\infty$ -- precisely our limit of interest) the authors of \cite{ven} sum up the amplitude
to all loop orders to a unitary eikonal form.  The large $R$ or large
energy limit localizes strings in the transverse directions and
reveals classical behaviour, much as the eikonal treatment in quantum
mechanics (or optics) reveals semiclassical particle (or ray)
behavior.

Note that $A(s,b)$ is dimensionless for any $D$.  It determines the
annihilation probability $P(b)$ via
\begin{equation}\label{Probb}
P(b)=\frac{1}{v}\text{Im}(A(s,b))\end{equation} 
with $\text{Im}A(s,b)$ as in
equation (\ref{im}) and $v$ the velocity of the colliding strings in
their center of mass frame.  This prescription can be shown to satisfy
the usual unitarity conditions in the large $s$ limit
\cite{adachi,cott}.

The quantity $\Delta x^2 \equiv 4Y \alpha' = 4 \alpha' \log
(R^2/\alpha')$ appearing in (\ref{im}) is interpreted as the quantum
thickness of the string.  It measures the fluctuations about the
classical straight string configuration. The fact that it increases
logarithmically with the string's length reflects that it is
energetically less costly to excite oscillators on a long
string. Similar string spreading effects occur in high energy
collisions and for strings falling into black holes
\cite{Susskind:1993aa}.  Note that this string spreading does not
include the effect of real (as opposed to virtual) oscillator
excitations as would be appropriate in the Hagedorn phase of a string
gas. In the Hagedorn phase wound strings are highly excited and their
spread in the transverse directions is comparable to the length of the
dimension they wrap \cite{turok}.\footnote{In the Hagedorn phase
strings perform a random walk in all directions. As their energy
scales with their length, their mean extent in all directions scales
as $\sqrt{E}$. This is the dependence of the winding number on
energy as we will see in the section on thermodynamics.} These
wiggly strings are very likely to intersect, leading to rapid
interactions which keep the strings in equilibrium.  But as the
universe expands and cools down the equilibrium phase becomes one of
pure radiation.  Then the oscillator excitations decay away and the
spread of the wound strings approaches $\Delta x$.  This justifies our
use of the amplitude (\ref{im}) if $b>\Delta x$.

It is useful to contrast the impact parameter picture to the more
standard method of obtaining a scattering probability. Typically one
derives a cross section $\sigma$ and the collision probability is
simply $n\sigma$ where $n$ is the number of targets per transverse
volume. If one has a collision probability in impact parameter space,
$P(b)$, then the scattering cross section is obtained via
\cite{peskin,taylor}
\begin{equation}\label{sig} \sigma =\frac{1}{n}\int d^{\perp}b \, n  P(b)\end{equation}
In other words, $n\sigma$ is an averaged probability in impact
parameter space. Most frequently it is assumed that the targets are uniformly distributed ($n =$
constant) and one obtains
\begin{equation}\label{non} \sigma = \int d^{\perp}b \,  P(b)\end{equation}
In the case of the optical theorem, for example, one can immediately derive
$\sigma=\frac{1}{v s}\text{Im} A(s,t=0)$ using equations (\ref{non}), (\ref{Probb}) and
(\ref{transf}). It is the $n =$ constant assumption that we are willing to relax here. 
There are two ways in which it can be justified.  First, if the targets are dense and uniform as in collider
experiments. A test particle in that case will interact with targets
at all impact parameters so one can integrate as in
(\ref{non}). Second, if the time between collisions is much smaller
than the total time over which collisions take place. Then the test
particle is given enough time to interact with targets at all impact
parameters (assuming each collision is at a random impact parameter)
and the averaging over impact parameters is essentially a time
average.

But if the winding modes in a string gas are dilute, with a mean
separation much larger than their thickness, the dense target
assumption above does not apply. It could still be that, since the
strings move in a compact space, they collide repeatedly with each
other and a time average is appropriate. It then becomes a matter of
timescales.  We need to compare the mean time between collisions with
the recollapse time, the time required for winding modes to pull the
universe back to a small-radius regime where winding modes are no
longer dilute.  An additional effect which must be taken into account
is that the string coupling is time dependent. This could also
invalidate the use of a time-averaged cross section.

We thus have to develop a model for the distribution of interactions
over impact parameters. We will return to this in the next section
after we set up the rest of the dynamics.

\section{Equations of motion}

In this section we write down coupled dilaton-gravity and Boltzmann
equations for the matter degrees of freedom. For further details on
the thermodynamic phases and energy conservation see \cite{winds,cyc}
and references therein.

The general setup is similar to \cite{winds}, except that instead of
an anisotropic universe we consider $d$ growing dimensions all with
the same radius, and hold the remaining $9-d$ dimensions frozen at the
self-dual radius. By removing the randomness in the choice of initial
radii present in \cite{winds} we can see more clearly the dependence
of the winding annihilations on the number of growing dimensions.

We consider type IIA string theory with a flat FRW metric on a torus
for the $d=D-1$ growing dimensions,
\begin{equation}
ds^2=-dt^2+\alpha' e^{2\lambda(t)} {\textstyle \sum_i} dx_i^2 \hspace{6mm} 0\leq x_i \leq 2\pi
\end{equation}
and a homogeneous shifted dilaton $\varphi(t)$. From now on we set $\alpha'=1$.
When the metric and dilaton are coupled to matter, the equations of motion are
\begin{equation}\label{eqmot}\begin{aligned}
\ddot{\varphi}&=\frac{1}{2}(\dot{\varphi}^2+d\dot{\lambda}^2)\\
\ddot{\lambda}&=\dot{\varphi}\dot{\lambda}+\frac{1}{8\pi^2}e^{\varphi}P
\end{aligned}\end{equation}
and the Hamiltonian constraint (Friedmann equation) is
\begin{equation}\label{ham}
E=(2\pi)^2 e^{-\varphi}(\dot{\varphi}^2-d\dot{\lambda}^2)
\end{equation}
Here $E$ is the total energy in the string gas and $P$ is the pressure
(times the volume) in $d$ dimensions. 

\subsection{Matter content and Boltzmann equations}
The background equations of motion are coupled to phenomenological
Boltzmann equations that govern the evolution of matter. We model
matter with three species:

$\bullet$ Winding modes that evolve according to
\begin{equation}\label{boltzW}
\dot{W}=-\Gamma_W (W^2-\langle W\rangle^2)
\end{equation}
We specify the interaction rates $\Gamma$ and the thermal equilibrium
values $\langle \cdot \rangle$ below. The total energy in winding and
anti-winding modes is $E_W=2dWe^{\lambda}$ and their contribution to
the pressure is $P_W=-2We^{\lambda}$.

$\bullet$ Radiation, or pure Kaluza-Klein modes, evolving according to
\begin{equation}\label{boltzK}
\dot{K}=-\Gamma_K (K^2-\langle K\rangle^2)
\end{equation}
The energy in KK and anti-KK modes is $E_K=2dKe^{-\lambda}$ and their
pressure is $P_K=2Ke^{-\lambda}$.

$\bullet$ Finally we include string oscillators, or massive string
modes, as pressureless matter.  The oscillator modes fill up the
energy budget via $E_{\text{osc}}=E-(E_W+E_k)$. We do not need a
Boltzmann equation for these modes since the dilaton-gravity equations
of motion automatically conserve energy, $dE=-PdV$.

\subsection{Thermodynamic phases and interaction rates}
Near the self-dual radius the gas of strings is in a high density
Hagedorn phase. The thermodynamics of this phase has been studied in
\cite{deo1,deo2}.  The quantities of interest here are the equilibrium
values of the winding and KK numbers\footnote{These values are derived
under the assumption that the microcanonical energy is split equally
amongst all dimensions, see \cite{winds}.}
\begin{equation}
\langle W \rangle=\frac{1}{12}\sqrt{\frac{E}{\pi}}e^{-\lambda}\hspace{6mm}
\langle K \rangle=\frac{1}{12}\sqrt{\frac{E}{\pi}}e^{\lambda}
\end{equation}
Since $E\gg 1$ most of the energy in the Hagedorn phase resides in
oscillator modes ($E_{\text{osc}}\simeq E-\sqrt{E}$).

As the volume of the universe grows and the energy density drops, the
equilibrium state should be one with only radiation.\footnote{See
\cite{bow,axe} for a detailed treatment of the conditions for
equilibrium between massive and massless modes in a string gas.} As
the condition for that transition we set
\begin{equation}
\frac{E}{V_d}\leq c_dT_H^{d+1}
\end{equation} 
with $T_H$ the Hagedorn temperature, $V_d=(2\pi)^de^{d\lambda}$ the
volume of the large dimensions, and $c_d$ a Stefan-Boltzmann constant
appropriate to the IIA gas of 128 massless Bose and Fermi degrees of
freedom in $d$ dimensions,
\begin{equation}
c_d=128\frac{2d!\zeta(d+1)}{(4\pi)^{d/2}\Gamma(d/2)}(2-1/2^d)
\end{equation}
In this phase the equilibrium values are
\begin{equation}
\langle W\rangle=0 \hspace{6mm}\langle K\rangle=\frac{1}{2d}Ee^{\lambda}
\end{equation}
That is, at equilibrium all the energy is in radiation (KK and anti-KK modes).

Now we need to specify the interaction rates entering the Boltzmann
equations.  Recall that for winding modes, with an impact parameter
$b$ in $D-4$ dimensions, the interaction probability is
$P(b)=\frac{1}{v}\text{Im}A(s,b)$ with Im$A(s,b)$ given in (\ref{im}).
For two wound strings moving in the $x_1$ direction and with opposite
winding along $x_2$, the right moving momenta are $p_{R1,2}=(E,\pm
Ev,\pm R/\alpha ')$ so $s_R=-(p_{R1}+p_{R2})^2=4E^2\simeq
(2R/\alpha')^2$ for slowly moving strings. Putting things together,
the interaction probability per unit time (per winding mode in the
direction of motion) can be written as
\begin{equation}\begin{aligned}\label{rate}
\Gamma_W &=\Gamma_0\times\Gamma_b\\
&\equiv\frac{\pi\alpha'}{4}\frac{\kappa_{10}^2}{V}\left(\frac{2R}{\alpha'}\right)^2\times\left[
\left(\frac{2\pi R}{(\pi\Delta x^2)^{\frac{1}{2}}}\right)^{D-4}e^{-\frac{b^2}{\Delta x^2}}\right]
\end{aligned}\end{equation}
with $V$ the total spatial volume of the 9-torus. In terms of our
variables, and with $\alpha'=1$, we have
$\frac{\kappa_{10}^2}{V}=\frac{1}{2(2\pi)}e^{\varphi}$ and
$R=e^{\lambda}$.  Note that $\Gamma_0$ is the interaction rate used in
\cite{winds}.

As explained earlier, an impact parameter representation is only
appropriate in the radiation phase, when the separation between
winding modes $r$ is larger than the string thickness $\Delta x$.
From the thermodynamic distributions of \cite{deo2} we can estimate
how the mean velocity $\bar{v}$ of a single winding mode depends on
$R$ and $E$ (see appendix A).\footnote{Even though we are working
off-equilibrium we consider the equilibrium velocities to be a good
approximation. In other words, we are assuming kinetic equilibrium and explore
the possibility of chemical equilibrium. To be precise, note that
$\bar{v}=v_{\text{rms}}=\sqrt{\langle v^2\rangle}$ is the root-mean squared velocity.} The mean time between collisions, or re-collision
time, is then $t_r \simeq \frac{r}{\bar{v}}$.  In practice, as we
numerically integrate the equations of motion, once we are in the
dilute regime we randomly choose an impact parameter $b$ on every
re-collision time. The impact parameter is chosen at random, from a
uniform distribution in the transverse $D - 4$ dimensions, up to the
maximum value $b = r$.  In effect, under the assumption of isotropy and even
distribution of winding modes, we treat their interactions as $d$-many copies
of a single interaction over a periodic lattice $r^d$ ($d=D-1$). We have now enough information to determine
$\Gamma_b$.

Another concern, raised earlier, is that in the dilute regime the
winding strings might not have time to collide before the universe
re-collapses to a dense Hagedorn phase.  In principle this could
happen even in $D = 4$. We test for this as follows.  Upon entering
the dilute regime we estimate the re-collision time $t_r$ and turn off
interactions, i.e.\ set $\Gamma=0$. If the negative pressure from the
frozen winding modes recollapses the universe in a time smaller than
$t_r$ it means that freezing the interactions was consistent, that is,
the winding modes truly had no time to collide. On the other hand, if
after time $t_r$ we are still in the dilute regime, then string
interactions must be taken into account.

Thus (\ref{rate}) provides our description of string interactions in
the dilute regime.  In the Hagedorn phase the strings have highly
excited oscillator modes which enhance the interaction rates since
more string is available. This was studied in \cite{winds} and it
amounts to inserting an overall factor of $\frac{16}{9}E$ in the
Boltzmann equations.

We also need to specify the interaction rate for KK modes. Since the
wavelength of these modes grows with $R$ a semiclassical impact
parameter picture at large $R$ is not appropriate.  Instead we should
average over impact parameters.  Since we already know the averaged
interaction rate $\Gamma_0$ for winding modes, by T-duality we can
take $\Gamma_K=\Gamma_0|_{\lambda\rightarrow -\lambda}$.

\subsection{Initial conditions}

We need to integrate the coupled equations (\ref{eqmot}),
(\ref{boltzW}), (\ref{boltzK}) subject to the constraint
(\ref{ham}). We need 6 initial conditions: $\varphi_0$,
$\dot{\varphi}_0$, $\lambda_0$, $\dot{\lambda}_0$, $W_0$ and $K_0$.
In this section we describe our method for sampling from the space of
initial conditions.  The basic idea is to fix the value of
$\dot{\varphi}$, scan over allowed values of $\lambda_0$ and
$\varphi_0$, and average over the values of $\dot{\lambda}_0$, $W_0$
and $K_0$ using a suitable probability distribution.

For effective supergravity to be valid and to ensure weak coupling we
take $\dot{\varphi}_0=-1$ and $\varphi_0 \lesssim -1$. Recall that the
dilaton evolves monotonically to weaker coupling with the absolute
value of its velocity decreasing \cite{winds}.

With $\dot{\varphi}_0$ and $\varphi_0$ fixed we consider the universe
at equilibrium with $\dot{\lambda}=0$ and $\lambda=0$ (at the self dual radius).  Calculating the
energy $E_{\text{eq}} = E|_{\dot{\lambda}=0}$ determines the
equilibrium number of winding and KK modes per dimension the universe would have at the self
dual radius,
\[
\langle W \rangle_{\text{sd}} = \langle K \rangle_{\text{sd}} = \frac{1}{12}\sqrt{\frac{E_{\text{eq}}}{\pi}}\,.
\]
We will use these values to set upper (lower) bounds for choosing
$W_0$ ($K_0$) below, once the volume fluctuates to larger values.

Next we choose the scale factor $\lambda_0$ of the
$d$-torus.\footnote{We consider the dynamics of $d$ dimensions, while
the remaining $9-d$ are kept frozen at the self dual radius,
$\lambda=0$ with $\dot{\lambda}=0$ for all times.}  We take
$\lambda_0 > 0$ since by T-duality we need not consider smaller
volumes. Having already determined $E_{\text{eq}}$, choosing
$\lambda_0$ fixes the energy density and thus the equilibrium
thermodynamic phase. 
The choice of $\lambda_0$ can be thought of as merely a choice of initial condition
but since we choose the winding and KK modes with respect to their values at
$\lambda=0$ (see below) it is more appropriate to interpret it as a
fluctuation from $\lambda=0$ to a larger
volume. Since we are choosing $\lambda_0$ at random we are then assuming the thermal
distribution to be flat. This is not an arbitrary choice. The entropy to
leading order in the Hagedorn phase is $S_0\simeq E/T_H$ and thus it costs no
entropy or energy for such a fluctuation. To next order the dependence of the
entropy on the radii $R=e^{\lambda}$ for $d$ spatial dimensions is \cite{Bassett,deo2} 
\[ S_1\simeq \log [1-\Gamma(2d)^{-1}(\eta E)^{2d-1}e^{-\eta E}]\]
with $\eta\sim 1/R^2$. Even though this contribution is very small for the ranges of
energies and radii we consider, it would be interesting to study more precisely the
effect that these corrections give to our scenario. 

The Hubble rate $\dot{\lambda}$ can also fluctuate away from zero, which is the
value that maximizes the entropy. In the
Hagedorn phase the entropy is given to a good approximation by
$S=E/T_H$. Thus $\dot{\lambda}_0$ is chosen randomly from the Gaussian
distribution
\begin{equation}\label{vel}e^S \propto e^{-\dot{\lambda}_0^2/(2\sigma_H^2)}\end{equation}
with $\sigma_H^2=\frac{T_He^{\varphi_0}}{2(2\pi)^2d}$. In the
radiation phase the entropy is
$S=\frac{d+1}{d}c_dV_d\left(\frac{E}{c_dV_d}\right)^{\frac{d}{d+1}}$. Using
(\ref{ham}), to leading order in $\dot{\lambda}_0$ we have the distribution
\begin{equation}
e^S \propto e^{-\dot{\lambda}_0^2/(2\sigma_r^2)}\end{equation}
with
\begin{equation}
\sigma_r^2=\left(\frac{\rho}{\rho_H}\right)^{\frac{1}{d+1}}\sigma_H^2,
\hspace{4mm}
\rho=\frac{E|_{\dot{\lambda}=0}}{V_d},\hspace{4mm}
\rho_H\equiv c_dT_H^{d+1}\end{equation}

It remains to choose the initial winding and KK numbers. Depending on
whether we are in the radiation or Hagedorn phase, the equilibrium
number of winding modes could be zero or not. Since we do not want to
begin with zero winding (we would not be testing the BV mechanism in
that case) the lowest value of $W$ we may pick is 0.5, our chosen
threshold between zero and non-zero winding. The furthest we can
fluctuate from equilibrium (the largest winding) is
$W_{\text{sd}}$. However it's possible that the volume is so large
that there isn't enough energy to support that much winding.  This
occurs if $W_{\text{sd}}>Ee^{-\lambda}/(2d)$. Putting everything
together, the initial winding number is chosen randomly in the range
\begin{equation} 
\left(\text{Max}\{0.5,\langle W\rangle
\},\text{Min}\{\langle W\rangle_{\text{sd}},\frac{E}{2d}e^{-\lambda}\} \right)
\end{equation}
In the Hagedorn phase the KK number can fluctuate between $\langle
K\rangle_{\text{sd}}$ and the equilibrium value at the given
$\lambda_0$, so we choose a value randomly in this range. In the
radiation phase, given $W_0$, we compute the energy in winding
$E_W=2dW_0e^{\lambda}$. The rest of the energy should be available to
radiation with the maximum KK number being
$K_{\text{max}}=(E-E_W)e^{\lambda}/(2d)$. Therefore $K_0$ is chosen
randomly in the range $(\langle
K\rangle_{\text{sd}},K_{\text{max}})$. If $K_{\text{max}}<\langle
K\rangle_{\text{sd}}$ we set $K_0=K_{\text{max}}$.\footnote{When
integrating the equations of motion we need to be careful not to
produce more radiation than energy conservation allows.  If at some
point $E=E_K+E_W$, that is all the oscillators decay, we set
$\langle K\rangle = (E-E_W)e^{\lambda}/(2d)$.}

Once initial conditions are fixed we integrate the equations of motion
until either the winding modes annihilate ($W<0.5$) or the
interactions freeze out, which we define as $\Gamma_0 W<0.1 H$. We use
the maximum rate $\Gamma_0$ instead of the total $\Gamma_W$ to allow
for the possibility that, depending on the randomly chosen value for
$b$, strings could interact even for $D>4$.

\section{Results}
With $\dot{\varphi}_0$ fixed at -1 we can scan initial conditions over
a two-dimensional lattice of points $(\lambda_0,\varphi_0)$. For each
lattice point we do 1000 runs to average over different choices of
$\dot\lambda_0$, $W_0$ and $K_0$.

\begin{figure}[htp]
\includegraphics*[scale=0.9,viewport=50 230 600 760,clip]{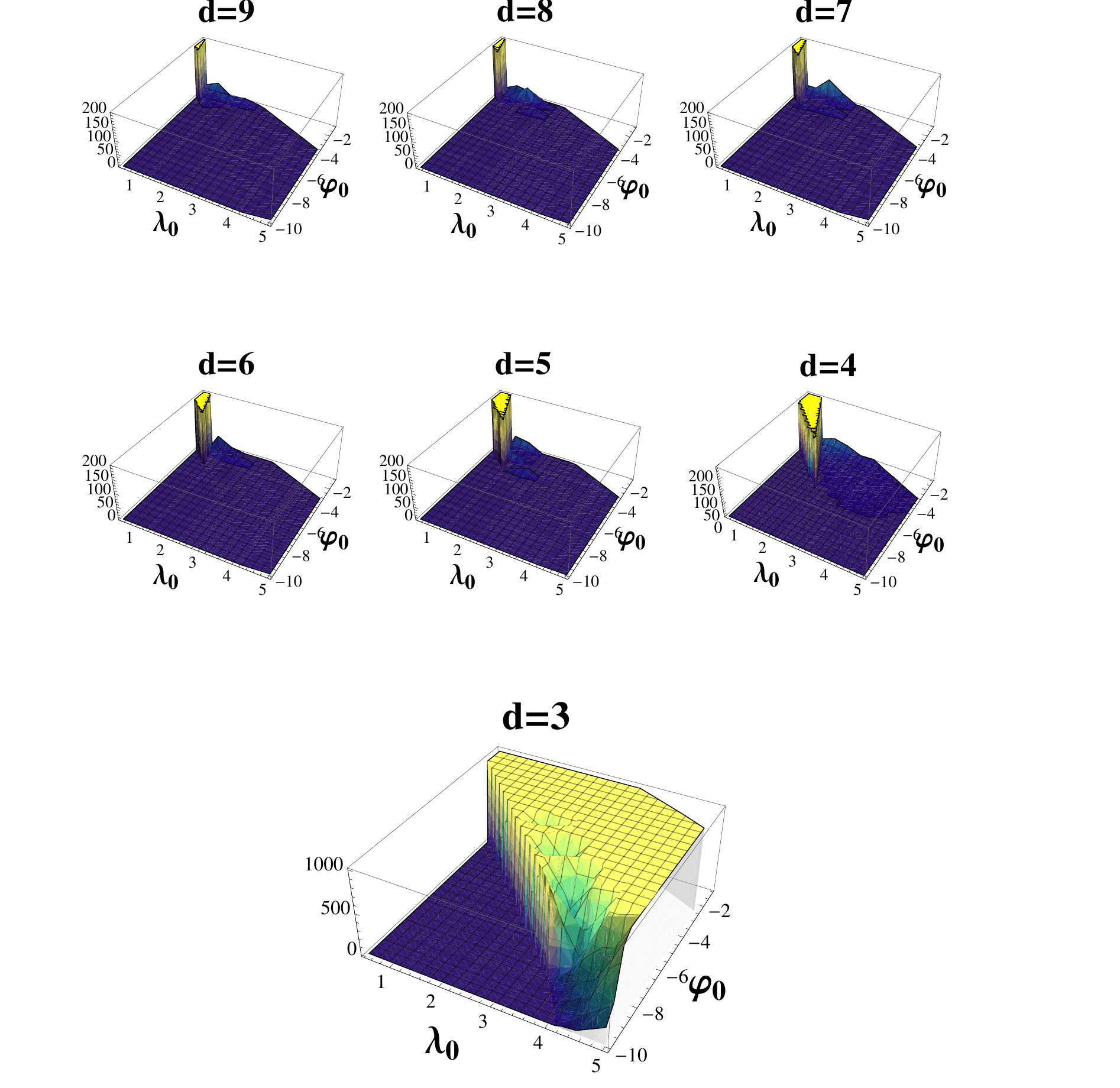}
\caption{
Number of cases decompactifying as a function of
$\varphi_0$ and $\lambda_0$ for different choices of growing dimensions
$d$. For $d=4,...,9$ the $z$-axis
is clipped at 200 to make the fewer decompactifying cases visible.}
\label{3d}
\end{figure}

\begin{figure}[htp]
\includegraphics[scale=0.8,viewport=30 300 600 780,clip]{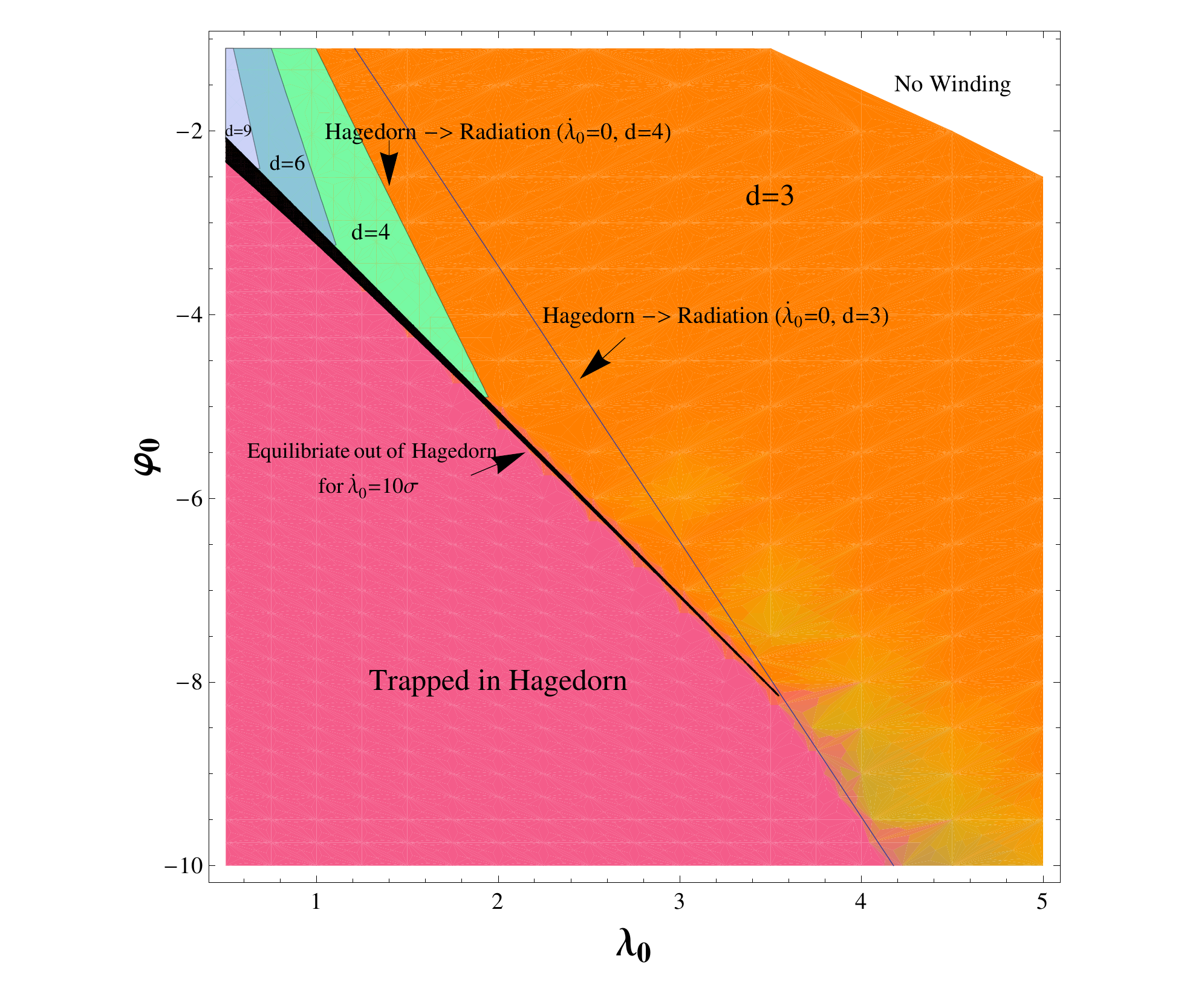}
\caption{A plot of the ($\lambda_0$, $\varphi_0$)
    plane contrasting the cases $d=3,\,4,\,6,\,9$.  If the initial
    equilibrium winding number in the Hagedorn phase is non-zero the
    system typically stays trapped in the Hagedorn phase, unless the
    initial Hubble rate is large and the initial winding number is small
    (thin dark region labeled `Equilibrate out of Hagedorn for
    $\dot{\lambda}_0=10\sigma$').  If the initial equilibrium winding
    number is zero in the Hagedorn phase then the universe typically
    decompactifies in any number of dimensions (regions in the upper left corner,
    labeled by dimension).  But if the universe
    begins in a radiation phase with a dilute gas of winding strings
    then only $d=3$ will decompactify (orange region to the right of the grey line).}
\label{map}
\end{figure}

The results for values of $d$ ranging from 9 to 3 are contrasted in
figures \ref{3d} and \ref{map}. For all values of $d$ we
find that equilibrium is maintained during the
Hagedorn phase. A consequence of this is that for a large range of
initial conditions, as long as the system starts out in the Hagedorn
phase, it will remain forever trapped there (see
figure \ref{map}). This is related to the limiting value of
$\lambda(t)$ as $t\rightarrow\infty$ in the solution to the equations
of motion (\ref{eqmot}) in thermal equilibrium \cite{cyc} and it is also the
same behaviour found in \cite{Bassett} at exact equilibrium.  When the
volume is large enough such that $\langle W \rangle \rightarrow
0$\footnote{In practice this would be $\langle W \rangle < 0.5$. We
round $W$ to zero and not $\langle W\rangle$.} yet the system is
still in the Hagedorn phase, then the universe decompactifies for any
$d$. This region gets narrower as $d$ increases as seen in figure
\ref{map}.  But for $d>3$ this is essentially the only region in
parameter space where decompactification occurs. If the system gets to
the radiation era, as the oscillators decay to massless modes and the
long strings start diluting, hardly any choice of initial conditions
leads to decompactification. For $d>3$ and outside the Hagedorn phase there
are few cases (of order $1\%$) in which the universe decompactifies. Those rare cases
have very small winding number and happen to have collisions at small impact parameters.

By contrast, for the $d=3$ case shown in figure \ref{3d}, we see that
even in the radiation phase long strings are able to annihilate. It is
interesting to note that the effects of large $\lambda$ enter in two
competing ways. First, due to the factor of $s\sim R^2=e^{2\lambda}$
in the amplitude, long strings interact more efficiently, even at weak
coupling. But at large $\lambda$ the effect of dilution is more
dramatic and strongly (exponentially) suppresses interactions for
$d>3$.

A comment on the choice $\dot{\varphi}_0=-1$. We could have considered
smaller (absolute) values for $\dot{\varphi}_0$, still valid within
supergravity. This can be compensated by a small (logarithmic) shift
in the initial dilaton to smaller values such that the initial energy
remains the same. The qualitative results should be unaltered.

\section{Summary and discussion}
The Brandenberger-Vafa mechanism relies on a classical
dimension-counting argument, namely that the worldvolumes of
one-dimensional objects will generically intersect in at most three
spatial dimensions. To see the mechanism at work one needs to be in a
regime where strings behave as semiclassical one-dimensional objects.
That is, one needs their length to be much larger than their effective
quantum thickness, and also their thickness to be much smaller than
the size of the transverse directions (the space is compact). We
realized these conditions in a simple isotropic setup where the length
of the string $R$ was the same as the size of the compactification
manifold. With oscillators excited, as in the Hagedorn phase, strings
have a significant spread in all directions and the classical picture
fails. But when oscillators decay, as in the radiation phase, the
thickness of strings grows as $\Delta x \sim \sqrt{\log R}$. Thus strings begin to
behave classically as $R/\sqrt{\log R}$ grows.  To model string interactions
in this regime we developed an impact parameter representation of the
string scattering amplitude.  This allowed us to show that in this
regime the BV mechanism indeed operates and favors decompactification
of three spatial dimensions.

To enter this regime we had to consider departures from equilibrium,
often large. Clearly in the radiation phase this is necessary since
the equilibrium number of winding strings is zero. In the Hagedorn
phase strings rapidly come to equilibrium and the pressure vanishes.
This means the universe tends to remain stuck in the Hagedorn phase,
and for some number of dimensions to decompactify a large fluctuation
is needed, either in the Hubble rate or in the initial volume, to send
the system to a regime where the equilibrium winding number is
zero. As the distribution (\ref{vel}) typically allows for only small
fluctuations in the Hubble rate, we had to consider large fluctuations
in the volume to realize the BV mechanism.  An important next step
would be to understand the likelihood of such a fluctuation taking
place in the early universe.

One shortcoming of our framework was that, even though we considered
a dilute gas of winding strings, we modeled the resulting pressure as
a homogeneous term in the gravity equations of motion.  This led us to
consider their back-reaction on spacetime in an all-or-nothing manner,
in which any amount of winding would oppose expansion while zero
winding would not. As far as testing the interactions and eventual
annihilation of strings, which was our focus, this shouldn't be a
concern. But a more detailed investigation of string gas cosmology
should address the issue of spatial inhomogeneity. Finally it would be
interesting to extend the analysis of this paper to the more general
context of M-theory, taking into account the effects of the full
$p$-brane spectrum.

\bigskip
\centerline{\bf Acknowledgements}
\noindent
DK is supported by U.S. National Science Foundation grant PHY-0855582 and
PSC-CUNY award \#60038-39-40, and is grateful to the Aspen Center for Physics
where this work was completed. SM is supported by an A.G.\ Leventis Foundation grant.

\bigskip
\goodbreak
\appendix

\section{Root-mean-square velocity of winding modes}
Thermodynamic quantities can be calculated using string distributions
derived by Deo, Jain and Tan \cite{deo1,deo2} in the microcanonical
ensemble. They show that the average number of strings with winding
charge vector $\mathbf{w}$, Kaluza-Klein charge vector $\mathbf{k}$
and energy $\epsilon$ on a $D$-torus with total energy $E$ is given by
\begin{equation}
D(\epsilon,\mathbf{w},\mathbf{k},E)=\frac{N}{\epsilon}u^De^{-u\mathbf{q}^TA^{-1}\mathbf{q}/4}
\end{equation}
where
\begin{equation*}\begin{aligned}
u&=\frac{E}{\epsilon(E-\epsilon)}\\
\mathbf{q}&=(\mathbf{w},\mathbf{k})\\
N&=\frac{(2\sqrt{\pi})^{-2D}}{\sqrt{\text{det}{A}}}\\
A&=\left(\begin{matrix}\frac{1}{4\pi^2R_i^2}\delta_{ij} & 0\\0 &
\frac{R^2_i}{4\pi^2}\delta_{ij}\end{matrix}\right)
\end{aligned}\end{equation*}
One can consider a unit winding mode, $\mathbf{w}_1=(1,0,...)$ along
one of the $d$ large dimensions, ($R_i=R$, $i=1,...,d$ and $R_j=1$,
$j=d+1,...,D$) and calculate the mean momentum squared of the string.
\begin{equation}\begin{aligned}
\langle k^2\rangle &= \frac{\int_0^Ed\epsilon \int d^D \mathbf{k}
\hspace{2mm}k^2\hspace{2mm}D(\epsilon,\mathbf{w}=\mathbf{w}_1,\mathbf{k},E)}
{\int_0^Ed\epsilon \int d^D\mathbf{k} \hspace{2mm} D(\epsilon,\mathbf{w}=\mathbf{w}_1,\mathbf{k},E)}\\[7pt]
&=\frac{dR^2}{2\pi^2}\frac{\int_0^E\frac{d\epsilon}{\epsilon}u^{D/2-1}e^{-u\pi R^2}}
{\int_0^E\frac{d\epsilon}{\epsilon}u^{D/2}e^{-u\pi R^2}}
\end{aligned}\end{equation}
Given the total energy $E$ and the radius $R$ the two integrals above can be
evaluated with saddle point methods or numerically. For a heavy winding mode
(large $R$) we have $\langle v^2\rangle\simeq \frac{\langle k^2\rangle}{R^4}$
and $v_{\text{rms}}=\sqrt{\langle v^2\rangle}$. 

\providecommand{\href}[2]{#2}\begingroup\raggedright\endgroup

\end{document}